\documentclass[preprint,superscriptaddress,nofootinbib,showkeys,onecolumn]{revtex4-1}

\usepackage{lmodern}
\usepackage[T1]{fontenc}
\usepackage[english,activeacute]{babel}
\usepackage{mathtools}
\usepackage{color} 

\usepackage{amssymb} 
\usepackage{dsfont} 

\usepackage{pbsi}

\usepackage[colorlinks,linkcolor=blue,citecolor=blue,urlcolor=blue]{hyperref}

\addto\extrasenglish{}
\addto\extrasenglish{}

\usepackage[symbol*]{footmisc}

\DefineFNsymbolsTM{otherfnsymbols}{%
  \textdagger *
   * *
  }%

\setfnsymbol{otherfnsymbols}

\begin{document}


\newcommand{\dd}{\mbox{d}}
\newcommand{\ii}{\mbox{i}}
\newcommand{\ddd}{\mbox{\scriptsize{d}}}
\newcommand{\iii}{\mbox{\scriptsize{i}}}

\newcommand{\nD}{n_{\mbox{\tiny{D}}}}
\newcommand{\nA}{n_{\mbox{\tiny{A}}}}
\newcommand{\nDss}{n_{\mbox{\tiny{D}}}^{\mbox{\tiny SS}}}
\newcommand{\nAss}{n_{\mbox{\tiny{A}}}^{\mbox{\tiny SS}}}

\newcommand{\nm}{\mbox{ nm}^{-3}}

\newcommand{\kET}{k_{\mbox{\tiny{ET}}}}
\newcommand{\gammaDrad}{\gamma_{\mbox{\tiny{D}}}^{\mbox{\scriptsize{r}}}}
\newcommand{\gammaArad}{\gamma_{\mbox{\tiny{A}}}^{\mbox{\scriptsize{r}}}}

\newcommand{\gammaorad}{\gamma_{0}^{\mbox{\scriptsize{r}}}}
\newcommand{\gammaonrad}{\gamma_{0}^{\mbox{\scriptsize{nr}}}}

\newcommand{\FET}{F_{\mbox{\tiny{ET}}}}
\newcommand{\Frad}{F^{\mbox{\scriptsize{r}}}}
\newcommand{\Fnrad}{F^{\mbox{\scriptsize{nr}}}}
\newcommand{\FDrad}{F_{\mbox{\tiny{D}}}^{\mbox{\scriptsize{r}}}}
\newcommand{\FArad}{F_{\mbox{\tiny{A}}}^{\mbox{\scriptsize{r}}}}

\newcommand{\omegaDem}{\omega_{\mbox{\tiny{D}}}^{\mbox{\scriptsize{em}}}}
\newcommand{\omegaDab}{\omega_{\mbox{\tiny{D}}}^{\mbox{\scriptsize{ab}}}}
\newcommand{\omegaAem}{\omega_{\mbox{\tiny{A}}}^{\mbox{\scriptsize{em}}}}
\newcommand{\omegaAab}{\omega_{\mbox{\tiny{A}}}^{\mbox{\scriptsize{ab}}}}

\title{Theory of Energy Transfer in Organic Nanocrystals}

\author{R. S\'aez-Bl\'azquez} \affiliation{Departamento de
F\'isica Te\'orica de la Materia Condensada and Condensed Matter
Physics Center (IFIMAC), Universidad Aut\'onoma de Madrid, E-28049
Madrid, Spain}

\author{J. Feist}
\affiliation{Departamento de F\'isica Te\'orica de la Materia
Condensada and Condensed Matter Physics Center (IFIMAC),
Universidad Aut\'onoma de Madrid, E-28049 Madrid, Spain}

\author{F. J. Garc\'ia-Vidal}
\email{fj.garcia@uam.es} \affiliation{Departamento de F\'isica
Te\'orica de la Materia Condensada and Condensed Matter Physics
Center (IFIMAC), Universidad Aut\'onoma de Madrid, E-28049 Madrid,
Spain} \affiliation{Donostia International Physics Center (DIPC),
E-20018 Donostia/San Sebasti\'an, Spain}

\author{A. I. Fern\'andez-Dom\'inguez}
\email{a.fernandez-dominguez@uam.es} \affiliation{Departamento de
F\'isica Te\'orica de la Materia Condensada and Condensed Matter
Physics Center (IFIMAC), Universidad Aut\'onoma de Madrid, E-28049
Madrid, Spain}

\begin{abstract}
Recent experiments have shown that highly efficient energy
transfer can take place in organic nanocrystals at extremely low
acceptor densities. This striking phenomenon has been ascribed to
the formation of exciton polaritons thanks to the photon
confinement provided by the crystal itself. We propose an
alternative theoretical model that accurately reproduces
fluorescence lifetime and spectrum measurements in these systems
without such an assumption. Our approach treats molecule-photon
interactions in the weak-coupling regime, and describes the donor
and acceptor population dynamics by means of rate equations with
parameters extracted from electromagnetic simulations. The
physical insight and predictive value of our model also enables us
to propose nanocrystal configurations in which acceptor emission
dominates the fluorescence spectrum at densities orders of
magnitude lower than the experimental ones.
\end{abstract}


\maketitle

\maketitle

Artificial light-harvesting systems have received much attention
lately. They are inspired by the pigment-protein antenna complexes
of natural photosynthesis, which convey the solar energy into the
reaction centers with efficiencies approaching
$100\%$~\cite{Scholes2011,Croce2014}. Mimicking bacterial and
plant photosynthetic units~\cite{Ziessel2011}, which present a
large number of antennas per reaction center~\cite{Mirkovic2017},
these artificial structures aim for high transfer efficiencies at
low acceptor/donor ratios~\cite{Roger2008}. Several experimental
configurations have been explored in this context, including
dendrimers and multiporphyrin arrays~\cite{Zhang2014, Choi2004},
multilayer polymer films~\cite{List2000, Kim2001}, and other
heterostructures~\cite{Zhang2014,Locritani2014} and supramolecular
compounds~\cite{Zhang2016,Li2017,Guo2018,Li2020}. In a recent
experiment carried out by Chen and coworkers~\cite{Chen2016},
$95\%$ transfer efficiency was reported in nanocrystals of
difluoroboron chromophores, in which aggregation-induced emission
did not occur~\cite{Hu2020} and with relative acceptor densities
of the order of $10^{-3}$. More surprisingly, the measured
fluorescence spectra showed fingerprints of energy transfer in
experimental samples with densities as small as $5\cdot10^{-5}$.

Concurrently with these advances in artificial light-harvesting,
research interests have also focused on polaritons as a means to
enhance energy transfer.
Experimental~\cite{Coles2014,Zhong2016,Zhong2017} and
theoretical~\cite{Du2018,Garcia-Vidal2017,SaezBlazquez2018}
reports have shown that the coherent and delocalized character of
these hybrid light-matter states makes it possible to increase the
spatial range and temporal rate of energy transfer processes.
Strong coupling between photons and molecular excitons has been
also investigated in photosynthetic
complexes~\cite{Coles2017,SaezBlazquez2019,Lishchuk2019}. Very
recently, it has been argued~\cite{Chen2019} that the formation of
polaritons is the underlying mechanism behind the efficient energy
transfer reported in organic nanocrystals at extremely low
acceptor densities. In this theory, the nanocrystal itself would
act as an optical cavity, providing the photon confinement
required for strong coupling.

In this Letter, we present a theoretical model for the process of
energy transfer in organic nanocrystals. Our approach depicts
photon-molecule interactions in the weak-coupling regime, and
accounts for the population dynamics of donor and acceptor
ensembles by means of coupled rate equations. These are
parameterized by radiative and nonradiative Purcell factor
simulations based on numerical solutions of Maxwell's Equations.
Without the need of any fitting procedure, our model accurately
describes the experimental results in Ref.~\onlinecite{Chen2016}.
Fluorescence spectra and donor lifetime measurements are
reproduced systematically by means of steady-state and transient
calculations for acceptor/donor ratios two orders of magnitude
apart. Our findings reveal that the combination of the short-range
F\"orster mechanism~\cite{Forster1959,Lakowicz1999} and the
inherently large donor densities in organic
nanocrystals~\cite{Jiang2018} makes energy transfer highly
efficient even in conditions of extremely low acceptor
concentrations. Contrary to what has been proposed in
Ref.~\onlinecite{Chen2019}, we show that there is no significant
photon confinement in these systems, which prevents the formation
of exciton polaritons. Finally, we employ our model to investigate
if the process of energy transfer could be modified by introducing
the samples in an optical cavity. We find that while its
efficiency cannot be tailored this way, it is possible to design
crystal-cavity configurations in which acceptor emission governs
the fluorescence spectrum at relative concentrations as small as
$10^{-5}$.

\begin{figure}[!t]
\centering
\includegraphics[width=0.5\linewidth]{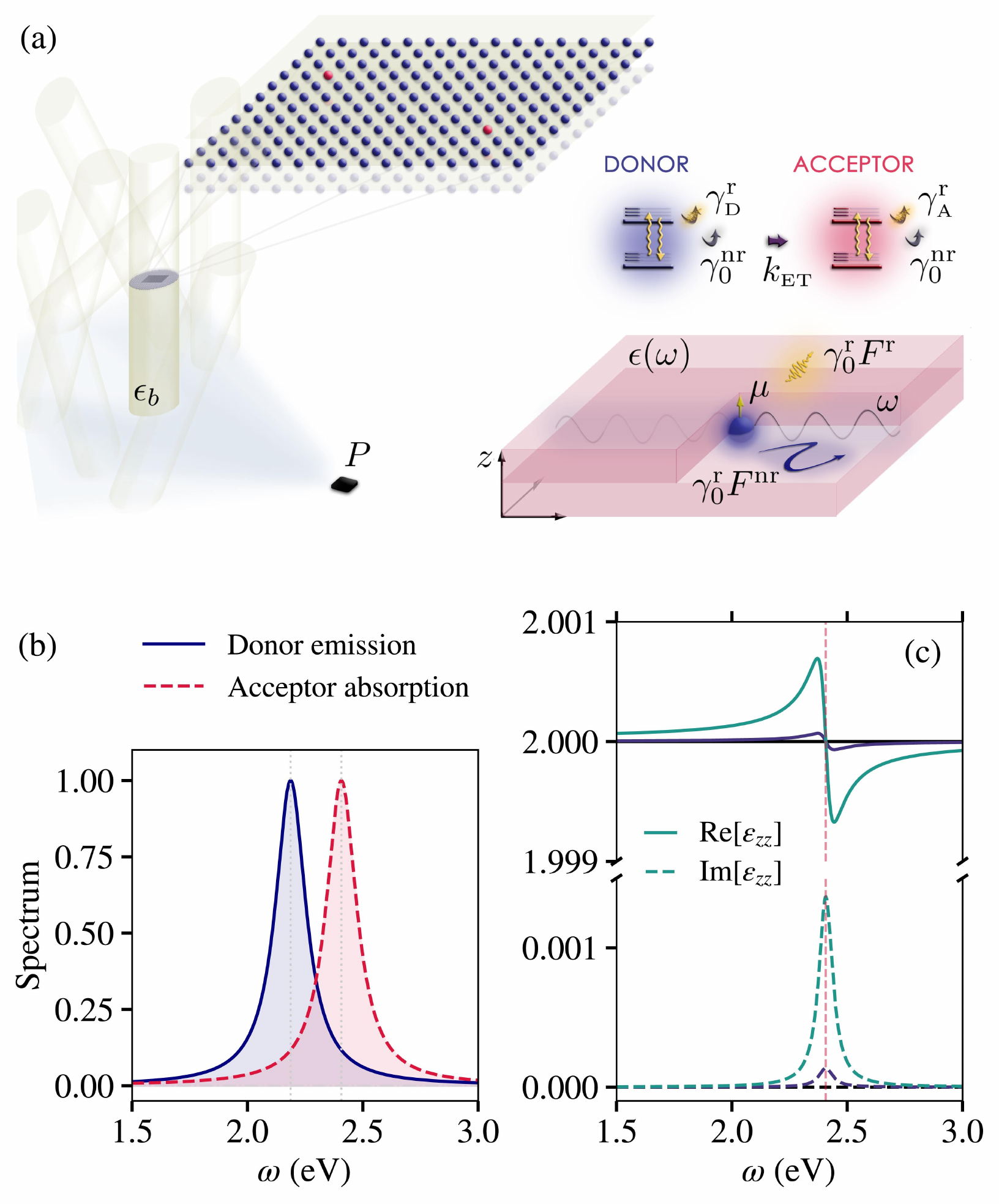}
\caption{(a) Scheme of the organic nanocrystal system, with the
relevant parameters characterizing our modelling. (b) Donor
emission (continuous blue line) and acceptor absorption (dashed
red line) spectra. (c) Real (continuous lines) and imaginary
(dashed lines) parts of the $zz$-component of the permittivity for
two different acceptor densities: $n=10^{-5} {\rm nm}^{-3}$ (blue)
and $n=10^{-4} {\rm nm}^{-3}$ (green).} \label{fig:Model}
\end{figure}

The first ingredient in our approach is the introduction of the
two-level system model for the donor and acceptor molecules. Both
are sketched in \autoref{fig:Model}(a) in blue and red colors,
respectively. The samples in Ref.~\onlinecite{Chen2016} were
composed of BF$_2$bcz (donors) nanocrystals presenting different
(controlled) densities of BF$_2$cna (acceptors) impurities in
their structure. Mimicking the experimental data, we set their
emission frequencies to $\omegaDem = 2.187$ eV and $\omegaAem =
2.013$ eV, respectively. The acceptor absorption is centered at
$\omegaAab = 2.407$ eV, overlapping significantly with the donor
emission, see \autoref{fig:Model}(b). All the spectra are modelled
as Lorentzian-like profiles of the form
\begin{equation} \label{eq:Lorentzian}
S^{o}_\iota (\omega) = \frac{1}{\pi} \frac{\sigma}{(\omega -
\omega_\iota^{o})^2 + \sigma^2} \ ,
\end{equation}
where $\iota = $ D, A, and o = em, ab, stands for emission or
absorption, respectively. We set $\sigma$ = 0.08 eV to account for
the width of the measured spectra. Note that, for simplicity, we
do not consider here BF$_2$dan acceptor chromophores, which were
also studied in the experiment yielding similar results as
BF$_2$cna samples.

The description of the energy transfer mechanism through
electromagnetic simulations requires treating donor molecules as
independent dipolar point sources. Their orientation is set by the
crystalline structure (we assume they point along the
$z$-direction). The ensemble of acceptor molecules embedded in the
crystal structure is modelled through an effective dielectric
function. Assuming that acceptor and donor chromophores have the
same orientation, the nanocrystal diagonal permittivity has the
form $\epsilon (\omega)=[\epsilon_b,\epsilon_b, \epsilon_{zz}
(\omega)]$. The lossless bare crystal permittivity is set to
$\epsilon_b = 2$~\cite{Chen2019} and
\begin{equation}
\epsilon_{zz} (\omega) = \epsilon_b \frac{1 + 2 n \alpha (\omega)
/(3 \epsilon_0)}{1 - n \alpha (\omega) / (3 \epsilon_0)} \
,\label{eq:permittivity}
\end{equation}
is given by the Clausius-Mossotti relation~\cite{Urano1977} where
$n$ is the number of acceptor molecules per unit volume and
$\epsilon_0$ is the vacuum permittivity. The polarizability of a
single acceptor chromophores reads~\cite{LoudonBook}
\begin{equation}
\alpha (\omega) = \frac{\mu^2}{\hbar} \frac{2
\omegaAab}{{(\omegaAab)}^2 - (\omega + \ii \sigma)^2}
\,\label{eq:alpha}
\end{equation}
where $\mu$ is the molecular transition dipole moment.

Difluoroboron chromophores present a fluorescence lifetime $\tau =
5.5$ ns and a quantum yield $\Phi = 0.4$~\cite{Chen2017}. Using
these values, we can compute their radiative and nonradiative
decay rates, $\gammaorad = \Phi / \tau$ and $\gammaonrad = (1 -
\Phi)  / \tau$, as well as the acceptor dipole moment in
\autoref{eq:alpha}, {$\mu=\sqrt{3\pi c^3\epsilon_0 \gamma_{0}^{\rm
r} /(\omega_{\rm A}^{\rm em})^3}=0.12$~${\rm e}\cdot{\rm
nm}$}~\cite{Novotnybook}. \autoref{fig:Model} (c) plots the real
(continuous lines) and imaginary (dashed lines) parts of the
$zz$-component of the nanocrystal permittivity as a function of
frequency for two different acceptor concentrations: $n=10^{-5}
\nm$ (blue) and $n=10^{-4} \nm$ (green). We can observe that, as
expected, ${\rm Im}\{\epsilon_{zz}(\omega)\}$ is governed by a
maximum at $\omega=\omegaAab$, the center of the acceptor
absorption band, marked with a vertical dashed red line. Note that
the maximum absorption increases with the acceptor density $n$.

We use the effective medium permittivity in
\autoref{eq:permittivity} to compute the Purcell factor
experienced by donor molecules in cylindrically-shaped
nanocrystals. Their diameter and height are set to 800 nm and 5
$\mu$m, respectively, in accordance with the dimensions of the
experimental samples. By numerically solving Maxwell's equations
using the finite element module of the commercial software COSMOL
Multiphysics, we can compute the fraction of the power emitted by
the donor molecules that is radiated into the far-field and
absorbed by the acceptor chromophore ensemble. Normalizing to the
power emitted in free-space, we can obtain the radiative and
nonradiative Purcell factors, $\Frad$ and $\Fnrad$, respectively.
We fix the donor density to 1~${\rm nm}^{-3}$, the inverse of the
volume per molecule reported theoretically for BF$_2$bcz
crystals~\cite{Jiang2018}. This value is also in agreement with
the interlayer distance reported experimentally. In order to link
the nonradiative Purcell factor with the energy transfer rate in
the systems, we exclude a cylindrical volume of 3 nm$^3$ around
the donor molecule in the calculation of $\Fnrad$. This way, we
account for the fact that molecular excitons are delocalized among
$\sim 10$ chromophores in BF$_2$bcz crystals~\cite{Jiang2018}.

\begin{figure}[!t]
\centering
\includegraphics[width=0.5\linewidth]{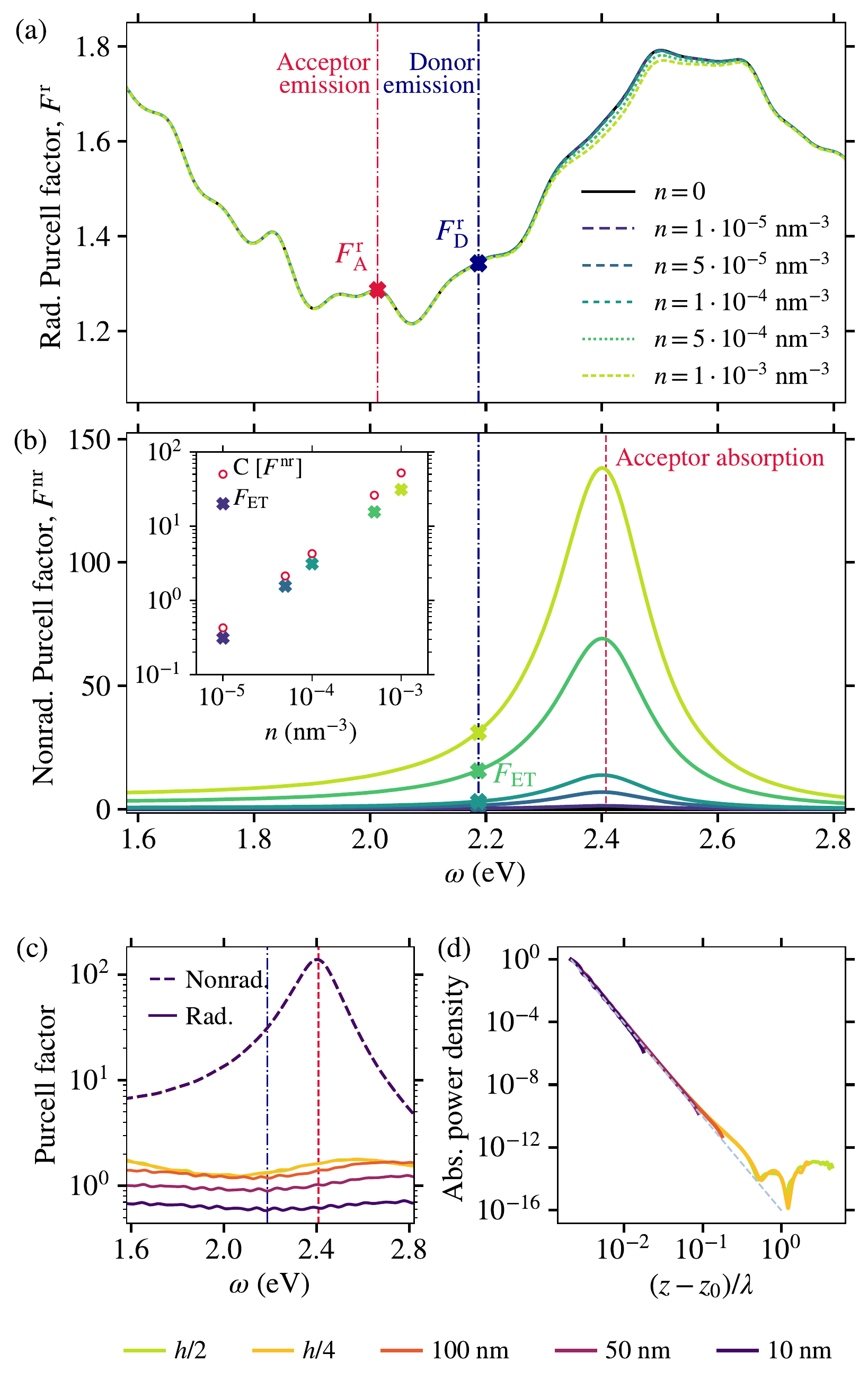}
\caption{(a-b) Radiative (a) and nonradiative (b) Purcell spectra
for different values of the acceptor density $n$. Vertical
dash-dotted lines indicate the donor (blue) and acceptor (red)
emission frequencies, and the dashed red line marks the acceptor
absorption frequency. The inset plots the normalized transfer rate
versus the acceptor density $n$ calculated as $\Fnrad(\omegaDem)$
and incorporating the whole donor emission band. (c) Absorbed
power density as a function of the distance from the donor,
$(z-z_0)$, normalized to the donor emission wavelength
$\lambda=2\pi c/\omegaDem$. A $1/(z-z_0)^6$ dependence is rendered
in grey dashed line. (d) Radiative (solid lines) and nonradiative
(dashed lines) Purcell spectra donor positions.}
\label{fig:PurcellFactors}
\end{figure}

\autoref{fig:PurcellFactors}(a) and (b) show radiative and
nonradiative Purcell spectra, respectively, for different acceptor
densities $n$. In these calculations, we have considered a single
donor chromophore placed in the center of the nanocrystal. We can
observe that $\Frad$ hardly varies when increasing the acceptor
concentration, and its value is always close to one. On the
contrary, $\Fnrad$ exhibits a peak centered at the acceptor
absorption frequency that, following ${\rm
Im}\{\epsilon_{zz}(\omega)\}$, grows with increasing $n$. As we
will show next, we can ascribe the normalized energy transfer rate
to the nonradiative Purcell factor evaluated at the donor emission
frequency, $\FET=\Fnrad(\omegaDem)$. This magnitude is plotted
against acceptor density in the inset of
\autoref{fig:PurcellFactors}(b) with coloured crosses. A more
rigorous description of the transfer rate incorporating the whole
donor emission band~\cite{Ringler2008}, $C[F^{\rm nr}]=\int
d\omega S^{\rm \tiny em}_{\rm D}(\omega)
\Fnrad(\omega)\gamma_0^{\rm r}(\omega)/\int d\omega S^{\rm \tiny
em}_{\rm D}(\omega)\gamma_0^{\rm r}(\omega)$, with $\gamma_0^{\rm
r}(\omega)=\mu^2\omega^3/3\pi c^3\epsilon_0$~\cite{Novotnybook},
is rendered in red empty circles. We can observe the agreement
between both sets of calculations, with slightly increasing
discrepancies at larger $n$. These results enable us to use
$\FET=\Fnrad(\omegaDem)$ in the following.

\autoref{fig:PurcellFactors}(c) explores the dependence of the
Purcell spectra on the position of the donor molecule, $z_0$. Five
different distances from the nanocrystal top face and along the
$z$-axis are considered. $\Fnrad$ (dashed lines) overlap for all
positions, from the crystal center to only 10 nm below its
boundary. This indicates that, as expected from the F\"orster
mechanism~\cite{Forster1959,Lakowicz1999}, each donor chromophore
transfers energy to those acceptor molecules that located only a
few nanometers away from it. The radiative Purcell factor (solid
lines) depends slightly on the donor position, ranging from 0.6 to
2. This is a clear indication that a strong photon confinement
does not take place in the system.

\autoref{fig:PurcellFactors}(d) plots the normalized absorbed
power density ($\FET$ per unit volume) as a function of the
vertical distance from the donor position, $(z-z_0)$, for the five
$z_0$ values in panel (c). The grey dashed line renders the $1/(z
- z_0)^6$ law characteristic of the F\"orster mechanism. We can
observe that the absorbed power density follows this trend for all
$z_0$, presenting only small deviations at the structure
boundaries. Importantly, we can infer that the proportional
relationship between $\FET$ and $n$ apparent in
\autoref{fig:PurcellFactors}(c) originates from the volume
integration of this inverse sixth-power dependence on the distance
of the absorbed power density.

We employ next our electromagnetic model to construct a
description of the donor and acceptor population dynamics in terms
of a system of rate equations. These must include the decay and
pumping channels experienced by each chromophore ensemble. The
radiative decay rates of donor and acceptor molecules within the
nanocrystal are computed from the Purcell factors evaluated at the
corresponding emission frequencies, $\gamma_{\rm D/A}^{\rm
r}=\gamma_0^{\rm r}\Frad(\omega_{\rm D/A}^{\rm em})$. Similarly,
as discussed above, the energy transfer rate at different acceptor
densities is given by $\kET = \gammaorad \FET=\gammaorad
\Fnrad(\omegaDem)$. The external excitation of the system is
accounted for through a pumping rate, $P$, acting only on the
donor molecules. Thus, we can write the rate equations as
\begin{equation}
\begin{split}
\frac{\dd \nD (t)}{\dd t} & = P  - \gammaonrad \nD (t) - \gammaDrad \nD (t) -  \kET \nD (t) \ ,
\\
\frac{\dd \nA (t)}{\dd t} & = \kET \nD (t)  - \gammaonrad \nA (t)
- \gammaArad \nA (t)  \ , \label{eq:RateEquations}
\end{split}
\end{equation}
where $\nD$ and $\nA$ stand for the donor and acceptor
populations, respectively. Notice that we have introduced an
additional term in the equations above, which describes the
nonradiative losses experienced by both chromophores due to their
relatively low intrinsic quantum yield.

The steady-state solution to \autoref{eq:RateEquations} describes
the continuous pumping of the donor molecules, in a similar way as
in a fluorescence measurement. This is obtained by imposing $\dd
n_{\rm D/A}(t)/ \dd t=0$, which yields the following constant
populations
\begin{equation}
\nDss = \frac{P}{\gammaonrad + \gammaDrad  +  \kET } \ ,
\hspace{4mm}
\nAss  = \frac{\kET}{\gammaonrad + \gammaArad} \nD^{\mbox{\tiny SS}}  \ .
\end{equation}
Once the donor and acceptor steady-state populations are known,
the fluorescence spectrum of the organic nanocrystal can be
written as
\begin{equation}
S (\omega) = \gammaDrad \nDss \omegaDem S_{\rm D}^{\rm \tiny em}
(\omega) +  \gammaArad \nAss \omegaAem S_{\rm A}^{\rm \tiny em}
(\omega) \ ,\label{eq:spectrum}
\end{equation}
where $S_{\rm D}^{\rm \tiny em}(\omega)$ and $S_{\rm A}^{\rm \tiny
em}(\omega)$ follow the Lorentzian profile given in
\autoref{eq:Lorentzian}, and we have used that the power radiated
by a single molecule scales as $\gamma_{\rm D/A}^{\rm
r}\omega_{\rm D/A}^{\rm em}$~\cite{Novotnybook}.

\begin{figure}[!t]
\centering
\includegraphics[width=0.5\linewidth]{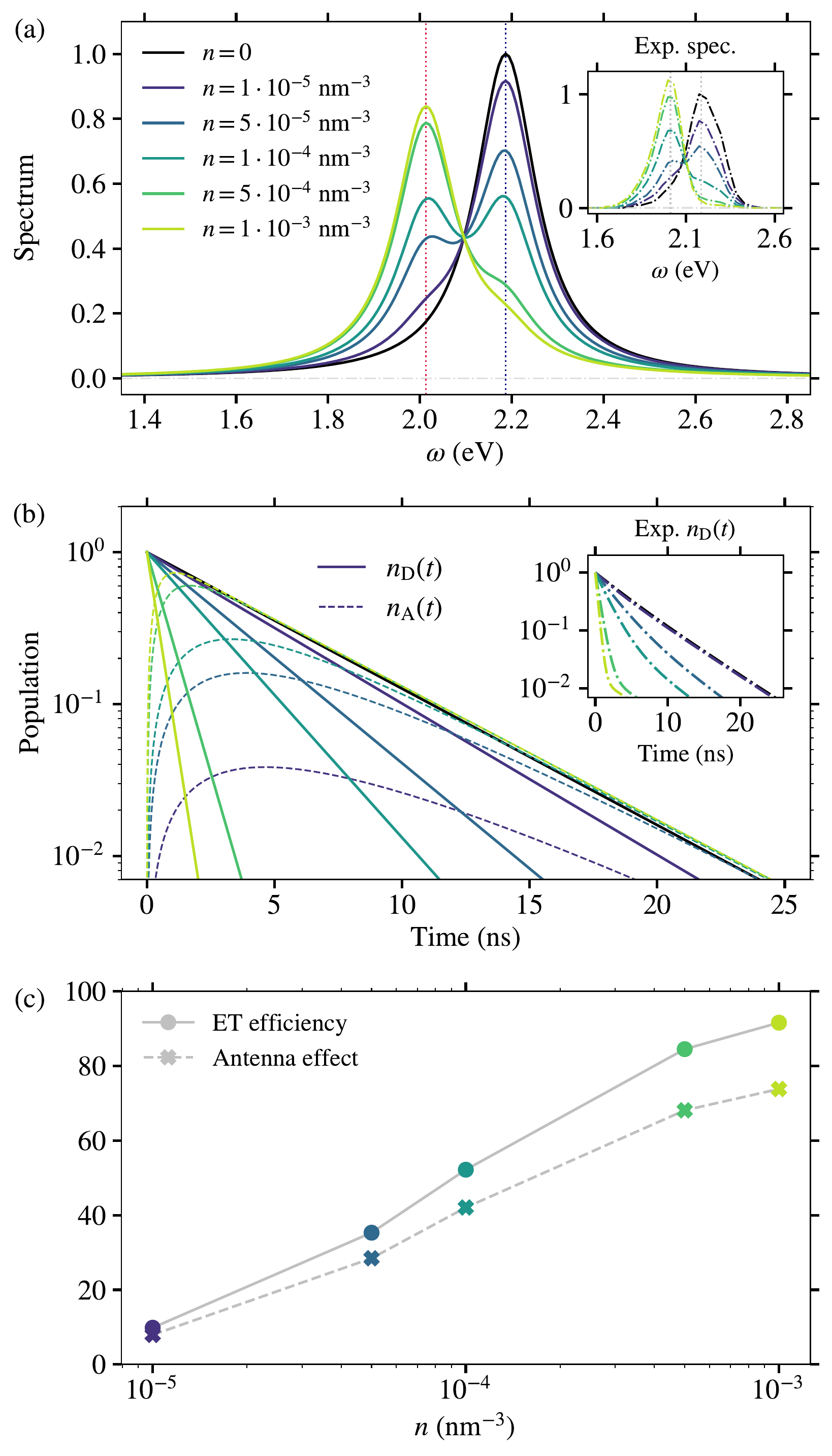}
\caption{(a) Theoretical fluorescence spectra for organic
nanocrystals  with different acceptor concentrations. Vertical
dotted lines indicate the donor (blue) and acceptor (red) emission
frequencies. (b) Time evolution of the donor (continuous lines)
and acceptor (dashed lines) populations for the same
configurations as in panel (a). The insets in (a) and (b) display
the experimental results in Ref.~\onlinecite{Chen2016}. (c) Energy
transfer efficiency (circles, continuous line) and antenna effect
(crosses, dashed line) as a function of $n$.} \label{fig:MimicExp}
\end{figure}

\autoref{fig:MimicExp}(a) displays fluorescence spectra obtained
from \autoref{eq:spectrum} for different values of the acceptor
density, ranging from $n = 10^{-5} \nm$ (purple) to $n = 10^{-3}
\nm$ (yellow). Within this window of acceptor/donor ratios
($10^{-5}$ to $10^{-3}$), the emission profile changes
qualitatively. The spectrum in absence of acceptor molecules
(black line) is shown as a reference (its maximum height is
normalized to 1). The donor emission dominating this configuration
decreases with larger acceptor density as a result of the energy
transfer mechanism. Concurrently, fluorescence from acceptor
molecules becomes more intense, and $S(\omega)$ is fully governed
by the acceptor Lorentzian at $n= 10^{-3} \nm$ . Importantly, the
acceptor contribution to the spectrum is clearly visible at
relative densities as low as $5\cdot 10^{-5}$. This evolution of
$S(\omega)$ with acceptor concentration is the main result of this
work, as it is in not only qualitative, but excellent quantitative
agreement with the experimental spectra in
Ref.~\onlinecite{Chen2016}. These are shown in the inset of
\autoref{fig:MimicExp}(a) using the same color code and
normalization as in the theoretical predictions to facilitate the
comparison. The predictive value of our model, which
systematically reproduces the dependence of the spectrum on $n$,
enables us to conclude that it captures all the relevant physical
mechanisms playing a role in the phenomenon of energy transfer in
organic nanocrystals. It also allows us to rule out the occurrence
of photon-molecule strong coupling and the formation of
polaritonic states in these systems.

We can employ \autoref{eq:RateEquations} to investigate the
population dynamics for an initial condition given by $\nD
(t=0)=1$, $\nA (t=0)=0$. By setting $P=0$,  we can mimic a
lifetime measurement configuration, in which donor molecules are
populated by an ultrafast laser pulse. In
\autoref{fig:MimicExp}(b), we plot the population transients for
the donor (continuous) and acceptor (dashed lines) molecules,
evaluated at the same densities as in panel (a). The population at
the donor molecules decays faster as $n$ increases, which is again
a clear signature of energy transfer to the acceptor chromophores.
These become significantly populated within less than 1 ns for $n
\simeq 10^{-4}\nm$, and $\nA>\nD$ within a few nanoseconds even
for lower values of the acceptor density. These findings also
match perfectly with the experimental observations by Chen and
coworkers~\cite{Chen2016}, shown as an inset. Here, and for
convenience, we plot the experimental multi-exponential fittings
to the measured data, rather than the measurements themselves.

To fully characterize the light-harvesting capabilities of the
organic nanocrystals, we compute next two physical magnitudes
widely employed in the experimental
literature~\cite{Li2017,Zhang2016}: the energy transfer efficiency
and the antenna effect. The former is usually defined as one minus
the ratio of the total fluorescence at $\omega=\omegaDem$ in the
absence (presence) of the acceptor ensemble. The latter is given
by the ratio of the acceptor contribution to the total
fluorescence at $\omega=\omegaAem$ under only donor and only
acceptor pumping conditions. Both magnitudes are shown in
\autoref{fig:MimicExp}(c) as a function of $n$. They follow a very
similar trend with lower values for the antenna effect at large
acceptor concentrations. For $n = 10^{-4} \nm$, the energy
transfer efficiency reaches 50 \%, and it amounts to 92 \% when $n
= 10^{-3} \nm$. This result is also in excellent agreement with
the experimental value (95 \%).

\begin{figure}[!t]
\centering
\includegraphics[width=0.5\linewidth]{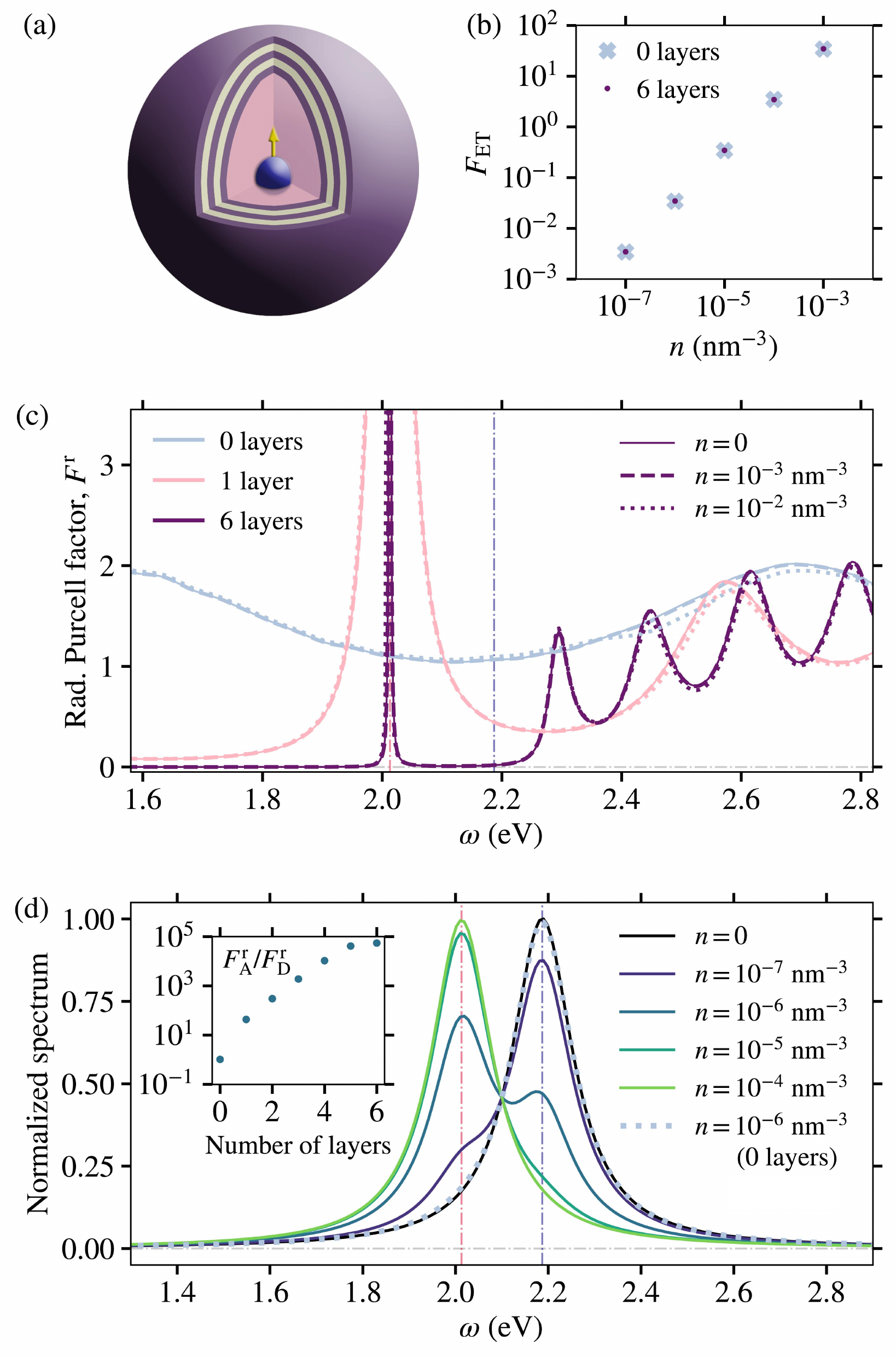}
\caption{(a) Sketch of a multilayered optical cavity embedding a
spherically-shaped organic nanocrystal. (b) Normalized energy
transfer rate versus acceptor density for bare (blue crosses) and
coated (purple dots) crystals. (c) Radiative Purcell spectra for
different number of cavity layers and acceptor densities. (d)
Fluorescence spectrum in 6-layered cavities and for low acceptor
concentrations. The blue dotted line corresponds to a bare
nanocrystal with  $n=10^{-6} \nm$, taken as reference. The inset
displays the ratio between the radiative Purcell factors evaluated
at the donor and acceptor emission frequencies versus the number
of cavity layers.} \label{fig:Layers}
\end{figure}

Finally, we apply our model beyond previous experimental
conditions and explore the impact that photon confinement has on
the fluorescence spectrum and the energy transfer efficiency in
organic nanocrystals. Our results reveal that it is absent in bare
structures, and therefore they must be placed within an optical
cavity for this purpose. \autoref{fig:Layers}(a) sketches the
simple system that we are investigating in the following: a 400 nm
radius spherical-shaped crystal is surrounded by a periodic
arrangement of 75 nm thick shells of two different materials.
These are set to the donor nanocrystal itself
($\epsilon=\epsilon_b=2$~\cite{Chen2019}), and a high refractive
index dielectric such as GaP ($\epsilon=9$~\cite{Cambiasso2017}).
As expected for the F\"orster mechanism, the optical coating does
not have any influence on the donor-acceptor transfer rates. This
is clearly shown by \autoref{fig:Layers}(b), which plots $\FET$ as
a function of $n$. Grey-blue crosses correspond to the uncoated
crystal, and purple dots to its embedding in a cavity comprising 6
periodic layers, each of them consisting in two shells of
different materials. These two sets of results do not only overlap
with each other, they also match perfectly the $\FET$ values in
the inset of \autoref{fig:PurcellFactors}(b), despite the fact
that they were calculated for different structure geometry and
dimensions.

Although the optical coating in \autoref{fig:Layers}(a) cannot
modify the energy transfer rate, it permits tailoring the
radiative Purcell spectrum for the system. This is apparent in
\autoref{fig:Layers}(c), which plots $\Frad$ versus frequency for
cavity-crystal configurations with three different number of
layers and acceptor/donor ratios. The cavity has been designed to
yield a sharp peak in the radiative Purcell at $\omega=\omegaAem$,
and a shallow dip at $\omega=\omegaDem$. We can observe that these
two spectral features are sensitive to the number of coating
layers but very robust to variations in $n$. By simple inspection
of \autoref{eq:spectrum}, we can infer that the radiative
enhancement of acceptor molecules and the inhibition of donor ones
effectively increases the weight of the acceptor contribution to
$S(\omega)$ at a fixed $n$. This is confirmed in
\autoref{fig:Layers}(d), which plots the normalized fluorescence
spectra of spherical crystals with extremely low acceptor
concentrations and surrounded by 6-layer cavities. It shows that
the emission by the acceptor molecules dominates $S(\omega)$ at
acceptor/donor ratios as small as $10^{-5}$, two orders of
magnitude lower than in \autoref{fig:MimicExp}(a). The influence
of photon confinement is also clearly illustrated by the
comparison of the two fluorescence spectra for $n=10^{-6}~\nm$.
The cavity transforms the single-peaked donor profile of the bare
structure into a doubled-peaked one, in which donor and acceptor
contributions have similar weights. The ratio between the
radiative Purcell factors for acceptor and donor chromophores in
the inset of \autoref{fig:Layers}(c) shows the strong dependence
of the cavity performance on the number of coating layers. These
results show that the fluorescence characteristics of organic
nanocrystals can be greatly modified through photon confinement,
despite the fact that energy transfer in these systems is
completely governed by the F\"orster mechanism.

In conclusion, we have presented a theoretical description of
energy transfer and fluorescence in organic nanocrsytals. It is
based on a rate equation treatment of donor and acceptor
population dynamics with parameters extracted from electromagnetic
simulations. The predictive value and the accuracy of our approach
has been demonstrated through a systematic comparison against
recent experimental results reporting high transfer efficiencies
at extremely low acceptor concentrations. Contrary to a previous
explanation of these results, we find that the crystal itself does
not provide a significant photon confinement, and therefore no
polaritonic effects take place in these systems. In contrast, it
is the extremely large donor density which makes F\"orster
transfer so efficient in these nanocrystals. Finally, we propose a
cavity-crystal configuration in which the acceptor channel
dominates the fluorescence intensity at concentrations orders of
magnitude lower than the experimental ones. We believe that our
theoretical model is a versatile, insightful and accessible tool
that may serve as guidance for the design and characterization of
fluorescence emission and energy transfer phenomena in complex
artificial photosynthetic structures.

This work has been funded by the Spanish Ministry for Science,
Innovation, and Universities - AEI grants RTI2018-099737-B-I00,
PCI2018-093145 (through the QuantERA program of the European
Commission), and MDM-2014-0377 (through the Mar\'ia de Maeztu
program for Units of Excellence in R\&D) and by the European
Research Council under Grant Agreement ERC-2016-STG-714870. It was
also supported by a 2019 Leonardo Grant for Researchers and
Cultural Creators, BBVA Foundation.

\bibliography{bibNanocrystals} 

\end{document}